\def\year{2018}\relax
\documentclass[letterpaper]{article} %DO NOT CHANGE THIS
\usepackage{aaai18}  %Required
\usepackage{times}  %Required
\usepackage{helvet}  %Required
\usepackage{courier}  %Required
\usepackage{url}  %Required
\usepackage{graphicx}  %Required

\usepackage{bm}
\usepackage{amsmath}
\usepackage{amssymb}
\usepackage[ruled, linesnumbered]{algorithm2e}
\DeclareMathOperator*{\argmax}{arg\,max}
\DeclareMathOperator{\sign}{sign}

\begin{document}
\title{Maximizing Activity in Ising Networks via the TAP Approximation}
\author{Christopher W. Lynn \\
Department of Physics \& Astronomy \\
University of Pennsylvania \\
Philadelphia, PA 19104, USA \\
\And Daniel D. Lee \\
Department of Electrical \& Systems Engineering \\
University of Pennsylvania \\
Philadelphia, PA 19104, USA}

\maketitle
\begin{abstract}
A wide array of complex biological, social, and physical systems have recently been shown to be quantitatively described by Ising models, which lie at the intersection of statistical physics and machine learning. Here, we study the fundamental question of how to optimize the state of a networked Ising system given a budget of external influence. In the continuous setting where one can tune the influence applied to each node, we propose a series of approximate gradient ascent algorithms based on the Plefka expansion, which generalizes the na\"{i}ve mean field and TAP approximations. In the discrete setting where one chooses a small set of influential nodes, the problem is equivalent to the famous influence maximization problem in social networks with an additional stochastic noise term. In this case, we provide sufficient conditions for when the objective is submodular, allowing a greedy algorithm to achieve an approximation ratio of $1-1/e$. Additionally, we compare the Ising-based algorithms with traditional influence maximization algorithms, demonstrating the practical importance of accurately modeling stochastic fluctuations in the system.
\end{abstract}

\section{Introduction}

The last 10 years have witnessed a dramatic increase in the use of maximum entropy models to describe a diverse range of real-world systems, including networks of neurons in the brain \cite{Schneidman-01,Ganmor-01}, flocks of birds in flight \cite{Bialek-01}, and humans interacting in social networks \cite{Lynn-03,Galam-02}, among an array of other social and biological applications \cite{Kapur-01,Phillips-01,Mora-01,Lezon-01}. Broadly speaking, the maximum entropy principle allows scientists to formalize the hypothesis that large-scale patterns in complex systems emerge organically from an aggregation of simple fine-scale interactions between individual elements \cite{Jaynes-01}. Indeed, intelligence itself, either naturally-occurring in the human brain and groups of animals \cite{Hillis-01} or artificially constructed in learning algorithms and autonomous systems \cite{Mataric-01,Namatame-01}, is increasingly viewed as an emergent phenomenon \cite{Levy-01}, the result of repeated underlying interactions between populations of smaller elements.

Given the wealth of real-world systems that are quantitatively described by maximum entropy models, it is of fundamental practical and scientific interest to understand how external influence affects the dynamics of these systems. Fortunately, all maximum entropy models are similar, if not formally equivalent, to the Ising model, which has roots in statistical physics \cite{Brush-01} and has a rich history in machine learning as a model of neural networks \cite{Coughlin-01}. The state of an Ising system is described by the average activity of its nodes. For populations of neurons in the brain, an active node represents a spiking neuron while inactivity represents silence. In the context of humans in social networks, node activity could represent the sending of an email or the consumption of online entertainment, while inactivity represents moments in which an individual does not perform an action. By applying external influence to a particular node, one can shift the average activity of that node. Furthermore, this targeted influence also has indirect effects on the rest of the system, mediated by the underlying network of interactions. For example, if an individual is incentivized to send more emails, this shift in behavior induces responses from her neighbors in the social network, resulting in increased activity in the population as a whole.

As a first step toward understanding how to control such complex systems, we study the problem of maximizing the total activity of an Ising network given a budget of external influence. This so-called \textit{Ising influence maximization} problem was originally proposed in the context of social networks \cite{Lynn-01}, where it has a clear practical interpretation: If a telephone company or an online service wants to maximize user activity, how should it distribute its limited marketing resources among its customers? However, we emphasize that the broader goal---to develop a unifying control theory for understanding the effects of external influence in complex systems---could prove to have other important applications, from guiding healthy brain development \cite{Goddard-01} and intervening to alleviate diseased brain states \cite{Goddard-02} to anticipating trends in financial markets \cite{Mantegna-01} and preventing viral epidemics \cite{Pastor-01}.

We divide our investigation into two settings: (i) the continuous setting where one can tune the influence applied to each node, and (ii) the discrete setting in which one forces activation upon a small set of influential nodes. In the continuous setting, we propose a gradient ascent algorithm and give novel conditions for when the objective is concave. We then present a series of increasingly-accurate approximation algorithms based on an advanced approximation technique known as the Plefka expansion \cite{Plefka-01}. The Plefka expansion generalizes the na\"{i}ve mean field and TAP approximations, and, in theory, can be extended to arbitrary order \cite{Yedidia-01}.

In the discrete setting, it was recently shown that Ising influence maximization is closely related to the famous influence maximization problem in social networks \cite{Kempe-01} with the addition of a natural stochastic noise term \cite{Lynn-02}. Here, we provide novel conditions for when the total activity of the system is submodular with respect to activated nodes. This result guarantees that a greedy algorithm achieves a $1-1/e$ approximation to the optimal choice of nodes. We compare our greedy algorithm with traditional influence maximization techniques, demonstrating the importance of accurately accounting for stochastic noise.

\subsection{Related Work}

Ising influence maximization was originally proposed in the context of human activity in social networks \cite{Lynn-01}. However, a recent surge in the use of Ising models to describe other biological and physical systems significantly expands the problem's applicability \cite{Stein-01}.

Ising influence maximization was originally studied in the continuous setting under the na\"{i}ve mean field approximation. Since the Plefka expansion generalizes the mean field approximation to increasing levels of accuracy, our work represents a principled improvement over existing techniques.

In the discrete setting, it was recently shown that Ising influence maximization is closely related to standard influence maximization \cite{Lynn-02}, which was first studied in the context of viral marketing \cite{Domingos-01}. Kempe et al. \cite{Kempe-01} proposed influence maximization as a discrete optimization problem and presented a greedy algorithm with approximation guarantees. Significant subsequent research has focused on developing efficient greedy and heuristic techniques \cite{Leskovec-01,Chen-02,Chen-03}. Here, we do not claim to provide improvements over these algorithms in the context of standard influence maximization. Instead, we focus on developing analogous techniques that are suitable for the Ising model.

\section{Ising Influence Maximization}

In the study of complex systems, if we look through a sufficiently small window in time, the actions of individual elements appear binary---either human $i$ sent an email ($\sigma_i = 1$) or she did not ($\sigma_i = -1$). The binary vector $\bm{\sigma} = \{\sigma_i\}\in \{\pm 1\}^n$ represents the activity of the entire system at a given moment in time, where $n$ is the size of the system.

Many complex systems in the biological and social sciences have recently been shown to be quantitatively described by the Ising model from statistical physics. The Ising model is defined by the Boltzmann distribution over activity vectors:
\begin{equation}
\label{Boltzmann}
P(\bm{\sigma}) = \frac{1}{Z}\exp\Bigg(\frac{1}{2}\sum_{i\neq j}J_{ij}\sigma_i\sigma_j + \sum_i b_i\sigma_i\Bigg),
\end{equation}
where $Z$ is a normalization constant. The parameters $J = \{J_{ij}\}$ define the network of interactions between elements and the parameters $\bm{b} = \{b_i\}$ represent individual biases, which can be altered by application of targeted external influence. For example, if $J$ defines the network of interactions in a population of email users, then $\bm{b}$ represents the intrinsic tendencies of users to send emails, which can be shifted by incentivizing or disincentivizing email use.

\subsection{Problem Statement}

The total average activity of a network with bias $\bm{b}$ is denoted $M(\bm{b}) = \sum_i\left<\sigma_i\right>$, where $\left<\cdot\right>$ denotes an average over the Boltzmann distribution (\ref{Boltzmann}). In what follows, we assume that the interactions $J$ and initial bias $\bm{b}^0$ are known. We note that this assumption is not restrictive since there exist an array of advanced techniques in machine learning \cite{Ackley-01} and statistical mechanics \cite{Aurell-01} for learning Ising parameters directly from observations of a system. 

We study the problem of maximizing the total activity $M$ with respect to an additional external influence $\bm{h}$, subject to the budget constraint $|\bm{h}|_p = \left(\sum_i | h_i |^p\right)^{1/p} \le H$, where $H$ is the budget of external influence. \\

\noindent \textbf{Problem 1 (Ising influence maximization).} Given an Ising system defined by $J$ and $\bm{b}^0$, and a budget $H$, find an optimal external influence $\bm{h}^*$ satisfying
\begin{equation}
\label{IIM}
\bm{h}^* = \argmax_{|\bm{h}|_p \le H} M(\bm{b}^0 + \bm{h}).
\end{equation}

\noindent We point out that the norm $p$ plays an important role. If $p = 1,2,3,\hdots$, then one is allowed to tune the influence on each node continuously. On the other hand, if $p = 0$, then $|\bm{h}|_0$ counts the number of non-zero elements in $\bm{h}$. In this case, one chooses a subset of $\lfloor H\rfloor$ nodes $\{i\}$ to activate with probability one by sending $\{h_i\} \rightarrow \infty$.

\section{The Plefka Expansion}

\noindent Since the Ising model has remained unsolved for all but a select number of special cases, tremendous interdisciplinary effort has focused on developing tractable approximation techniques. Here, we present an advanced approximation method known as the Plefka expansion \cite{Yedidia-01}. The Plefka expansion is not an approximation itself, but is rather a principled method for deriving a series of increasingly accurate approximations, the first two orders of which are the na\"{i}ve mean-field (MF) and Thouless-Anderson-Palmer (TAP) approximations. In subsequent sections, we will use the Plefka expansion to approximately solve the Ising influence maximization problem in (\ref{IIM}).

Calculations in the Ising model, such as the average activity $\left<\sigma_i\right>$, generally require summing over all $2^n$ binary activity vectors. To get around this exponential dependence on system size, the Plefka expansion provides a series of approximations based on the limit of weak interactions $|J_{ij}|\ll 1$. Each order $\alpha$ of the expansion generates a set of self-consistency equations $m_i = f_i^{(\alpha)}(\bm{m})$, where $m_i$ approximates the average activity $\left<\sigma_i\right>$. Thus, for any order $\alpha$ of the Plefka expansion, the intractable problem of computing the averages $\left<\sigma_i\right>$ is replaced by the manageable task of computing solutions to the corresponding self-consistency equations $\bm{m} = \bm{f}^{(\alpha)}(\bm{m})$. We point the interested reader to Appendix A for a detailed derivation of the Plefka expansion.

For a system with interactions $J$ and bias $\bm{b}$, the first order in the expansion yields the na\"{i}ve mean field approximation, summarized by the self-consistency equations
\begin{equation}
m_i = \tanh\Big[b_i + \sum_j J_{ij}m_j \Big] \triangleq f^{\text{MF}}_i(\bm{m}).
\end{equation}
The second order in the Plefka expansion yields the TAP approximation,
\begin{align}
m_i &= \tanh\Big[b_i + \sum_j J_{ij}m_j - m_i\sum_j J_{ij}^2(1-m_j^2) \Big] \nonumber \\
&\triangleq f^{\text{TAP}}_i(\bm{m}).
\end{align}
Higher-order approximations can be achieved by systematically including higher orders of $J$ in the argument of $\tanh[\cdot]$. In Appendix B, we present a derivation of the third-order approximation, denoted TAP3.

The standard approach for computing solutions to the self-consistency equations $\bm{m} = \bm{f}^{(\alpha)}(\bm{m})$ is to iteratively apply $\bm{f}^{(\alpha)}$ until convergence is reached:
\begin{equation}
\bm{m} \leftarrow (1-\gamma)\bm{m} + \gamma\bm{f}^{(\alpha)}(\bm{m}),
\end{equation}
where $\gamma\in [0,1]$ is the step size. The convergence of this procedure was rigorously examined in \cite{Bolthausen-01}. In practice, we find that $\gamma\sim 0.01$ yields rapid convergence for most systems up to the third-order approximation.

\section{The Continuous Setting}

\begin{figure*}
\includegraphics[width = \textwidth]{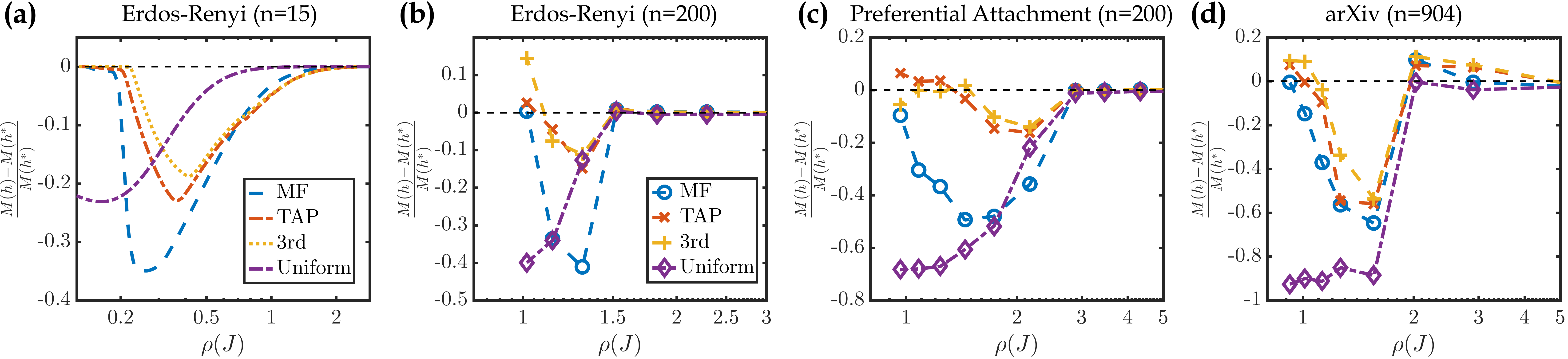}
\caption{\label{continuous} Performance of PGA for various orders of the Plefka expansion. (a) An Erd\"{o}s-R\'{e}nyi network with $n=15$ nodes and budget $H=1$. The total activity is calculated exactly using the Boltzmann distribution. (b) An Erd\"{o}s-R\'{e}nyi network with $n=200$ nodes and budget $H=10$. (c) A preferential attachment network with $n=200$ nodes and budget $H=10$. (d) A collaboration network of $n=904$ physicists on the arXiv and budget $H=20$. The total activities in (b-d) are estimated using Monte Carlo simulations. The benchmarks are PGA with the exact gradient for (a) and the gradient estimated using Monte Carlo simulations in (b-d).}
\end{figure*}

In this section, we study Ising influence maximization under a budget constraint $|\bm{h}|_p\le H$, where $p=1,2,3,\hdots$, yielding a continuous optimization problem where one can tune the external influence on each element in the system. We first present an exact gradient ascent algorithm and comment on its theoretical guarantees. We then demonstrate how the Plefka expansion can be used to approximate the gradient, yielding a series of increasingly accurate approximation algorithms.

\subsection{Projected Gradient Ascent}

We aim to maximize the total activity $M(\bm{b}^0 + \bm{h}) = \sum_i\left<\sigma_i\right>$ with respect to the external influence $\bm{h}$. Thus, a crucially important concept is the response function
\begin{equation}
\chi_{ij} = \frac{\partial \left<\sigma_i\right>}{\partial h_j} = \left<\sigma_i\sigma_j\right> - \left<\sigma_i\right>\left<\sigma_j\right>,
\end{equation}
which quantifies the change in the activity of node $i$ due to a shift in the influence on node $j$. The gradient of $M$ with respect to $\bm{h}$ can be succinctly written $\nabla_{\bm{h}}M = \chi^T\bm{1}$, where $\bm{1}$ is the $n$-vector of ones.

In Algorithm 1 we present a projected gradient ascent algorithm PGA. Starting at a feasible choice for the external influence $\bm{h}^{(0)}$, PGA steps along the gradient $\chi^T\bm{1}$ and then projects back down to the space of feasible solutions $|\bm{h}|_p\le H$. We note that for $p=1,2,3,\hdots$, the space of feasible solutions is convex, and hence the projection $\pi_{|\bm{h}|_p\le H}$ is well-defined and can be performed efficiently \cite{Duchi-01}.

\begin{algorithm}[t]
\KwIn{Ising system defined by $J$ and $\bm{b}^0$, budget $H$, norm $p$, and error $\epsilon$;}
\KwOut{External influence $\bm{h}$;}
\textbf{Initialize:} Choose $|\bm{h}^{(0)}|_p \le H$,  $k\leftarrow 0$; \\
\While{$\left|M(\bm{b}^0+\bm{h}^{(k)}) - M(\bm{b}^0+\bm{h}^{(k-1)})\right| >\epsilon$}{
Choose step size $\eta_k$;
$\bm{h}^{(k+1)} \leftarrow \pi_{|\bm{h}|_p\le H}\left[\bm{h}^{(k)} + \eta_k \chi(\bm{b}^0+\bm{h}^{(k)})^T\bm{1}\right]$\;
$k$++\;
}
$\bm{h}\leftarrow \bm{h}^{(k)}$\;
\caption{Projected Gradient Ascent (PGA)}
\label{PGA}
\end{algorithm}

\subsection{Conditions for Optimality}

The algorithm PGA efficiently converges to an $\epsilon$-approximation of a local maximum of $M$ in $O(1/\epsilon)$ iterations \cite{Nesterov-01}. However, this local maximum could be arbitrarily far from the globally optimal solution. Here, we present a novel sufficient condition for when PGA is guaranteed to converge to a global maximum of $M$, subject to the proof of a long-standing conjecture.\\ 

\noindent \textbf{Conjecture 2 \cite{Sylvester-01}.} Given an Ising system with non-negative interactions $J\ge 0$ and non-negative biases $\bm{b}\ge0$, the average activity of each node $\left<\sigma_i\right>$ is a concave function of the biases $\bm{b}$. \\

\noindent \textbf{Theorem 3.} If Conjecture 2 holds, then for any Ising system with non-negative interactions $J\ge 0$ and non-negative initial biases $\bm{b}^0\ge 0$, PGA converges to a global maximum of the total activity $M$. \\

\noindent \textit{Proof.} For Ising systems with positive couplings $J\ge 0$, the response function is non-negative $\{\chi_{ij}\} \ge 0$ \cite{Griffiths-03}. This implies two things: (i) at least one global maximum of $M(\bm{b})$ occurs in the non-negative orthant of $\bm{b}$, and (ii) if $\bm{b}^0\ge 0$, then $\bm{b}^0+\bm{h}^{(k)}$ will be non-negative at every iteration $k$ of PGA. If Conjecture 2 holds, then every local maximum in the non-negative orthant is a global maximum. Thus, PGA converges to a global maximum. \hfill $\square$ \\

We remark that Sylvester \cite{Sylvester-01} provides extensive experimental justification for Conjecture 2, and even proves Conjecture 2 in a number of limited cases. Additionally, we manually verified the veracity of Conjecture 2 in each of the experiments presented below. We also note that the sufficient conditions are plausible for many real-world scenarios. Positive interactions $J\ge 0$ imply that an action from one node will tend to induce an action from another node, a phenomenon that has been experimentally verified in small neuronal \cite{Schneidman-01} and social \cite{Lynn-03} networks. The more stringent condition is that $\bm{b}^0\ge$, implying that each element in the network prefers activity over inactivity.

\subsection{Approximating the Gradient via the Plefka Expansion}

Since PGA requires calculating the response function $\chi$ at each iteration, an exact implementation scales exponentially with the size of the system. Here we show that the Plefka expansion can be used to approximate $\chi$, yielding a series of efficient and increasingly-accurate gradient ascent algorithms.

Given a self-consistent approximation of the form $\bm{m} = \bm{f}^{(\alpha)}(\bm{m})$, where $\alpha$ denotes the order of the Plefka approximation, the response function is approximated by
\begin{equation}
\tilde{\chi}^{(\alpha)}_{ij} = \frac{\partial f^{(\alpha)}_i}{\partial h_j} + \sum_k \frac{\partial f^{(\alpha)}_i}{\partial m_k}\tilde{\chi}^{(\alpha)}_{kj}.
\end{equation}
For all orders $\alpha$ of the Plefka expansion, we point out that $\partial f^{(\alpha)}_i/\partial h_j = (1-m_i^2)\delta_{ij}$. Thus, defining $A_{ij} \triangleq (1-m_i^2)\delta_{ij}$, and denoting the Jacobian of $\bm{f}^{(\alpha)}$ by $Df^{(\alpha)}_{ij} \triangleq \partial f^{(\alpha)}_i/\partial m_j$, the response function takes the particularly simple form
\begin{equation}
\tilde{\chi}^{(\alpha)} = (I - Df^{(\alpha)})^{-1}A,
\end{equation}
where $I$ is the identity matrix. 

Thus, to approximate the gradient $\nabla_{\bm{h}} M \approx \tilde{\chi}^T\bm{1}$, one simply needs to calculate the Jacobian of $\bm{f}^{(\alpha)}$. Under the mean field approximation, the Jacobian takes the form $Df^{\text{MF}} = AJ$; and under the TAP approximation, we have
\begin{equation}
Df^{\text{TAP}}_{ij} = (1-m_i^2)\Big[J_{ij} + 2J_{ij}m_im_j - \delta_{ij}\sum_kJ_{ik}(1-m_k^2)\Big].
\end{equation}
We point the reader to Appendix B for a derivation of the third-order Jacobian.

\subsection{Experimental Evaluation}

In Fig. \ref{continuous}, we compare various orders of the Plefka approximation across a range of networks for the norm $p=1$. We also compare with the uniform influence $\bm{h}=H/n\bm{1}$ as a baseline. In Fig. \ref{continuous}(a), the network is small enough that we can calculate the exact optimal solution $\bm{h}^*$, while for Figs. \ref{continuous}(b-d), we approximate $\bm{h}^*$ by running costly Monte Carlo simulations to estimate the gradient at each iteration of PGA. Similarly, we calculate the total activity $M$ exactly using the Boltzmann distribution in Fig. \ref{continuous}(a), while in Figs. \ref{continuous}(b-d), we estimate $M$ using Monte Carlo simulations.

For each network, we assume that the interactions are symmetric $J=J^T$ with uniform weights and that the initial bias is zero $\bm{b}^0=0$. We then study the performance of the various algorithms across a range of interaction strengths, summarized by the spectral radius $\rho(J)$. For $\rho(J)\ll 1$, the network is dominated by randomness and all influence strategies have little affect on the total activity. On the other hand, for $\rho(J) \gg 1$, the elements interact strongly and any positive influence induces the entire network to become active. Thus, the interesting regime is the ``critical" region near $\rho(J)\approx 1$.

The striking success of the Plefka expansion is summarized by the fact that TAP and TAP3 consistently provide dramatic improvements over the na\"{i}ve mean field algorithm studied in \cite{Lynn-01}. Indeed, TAP and TAP3 consistently perform within 20\% of optimal (except for the arXiv network) and sometimes even outperform the Monte Carlo algorithm benchmark in Figs. \ref{continuous}(b-d). Furthermore, while the Monte Carlo algorithm takes $\sim10$ minutes to complete in a network of size $200$, PGA with the TAP and TAP3 approximations converges within $\sim 5$ seconds.

\section{Discrete Setting}

We now consider the discrete setting corresponding to a budget constraint of the form $|\bm{h}|_0\le H$. In this setting, one is allowed to apply infinite external influence to a set of $\lfloor H\rfloor$ nodes in the system, activating them with probability one; that is, one chooses a set of nodes $V=\{i\}$ for which we impose $\left<\sigma_i\right> = 1$ by taking $h_i\rightarrow +\infty$. We begin by presenting a greedy algorithm that selects the single node at each iteration that yields the largest increase in the total activity $M$. We then provide novel conditions for when $M$ is submodular in the selected nodes, which guarantees that our greedy algorithm is within $1-1/e$ of optimal. Finally, we comment on the relationship between (discrete) Ising influence maximization and traditional influence maximization in viral marketing, and we present experiments comparing our greedy algorithm with traditional techniques.

\subsection{A Greedy Algorithm}

We aim to maximize the total activity $M$ with respect to a set $V$ of activated nodes of size $|V|=H$ (assuming $H$ is integer). To eliminate confusion, we denote by $\mathcal{M}(V)$ the total activity of the system after activating the nodes in $V$, assuming that the couplings $J$ and initial bias $\bm{b}^0$ are already known.

Since there are $\binom{n}{H}\sim n^H$ possible choices for $V$, an exhaustive search for the optimal set is generally infeasible. On the other hand, we can simplify our search by looking at one node at a time and iteratively adding to $V$ the single node that increases $\mathcal{M}$ the most. This approximate greedy approach was made famous in traditional influence maximization in the context of viral marketing by Kempe et al. \cite{Kempe-01}. In Algorithm \ref{Greedy} we propose an analogous algorithm for computing the top $H$ influential nodes in an Ising system.

\begin{algorithm}[t]
\KwIn{Ising system defined by $J$ and $\bm{b}^0$, budget $H$;}
\KwOut{Set of $H$ influential nodes $V$;}
\textbf{Initialize:} $V^{(1)}\leftarrow \{\}$; \\
\For{$k = 1,\hdots,H$}{
\For{all nodes $i\in\{1,\hdots,n\}/V^{(k)}$}{
Calculate total activity $\mathcal{M}(V^{(k)}\cup \{i\})$\;
}
Choose node $i^* = \argmax_i \mathcal{M}(V^{(k)}\cup \{i\})$\;
Add $i^*$ to influential set $V^{(k+1)}\leftarrow V^{(k)}\cup \{i^*\}$\;
}
$V\leftarrow V^{(k+1)}$\;
\caption{Greedy algorithm for choosing top $H$ influential nodes in an Ising network (GI)}
\label{Greedy}
\end{algorithm}

\subsection{Theoretical Guarantee}

\begin{figure*}[t]
\centering
\includegraphics[width = .9\textwidth]{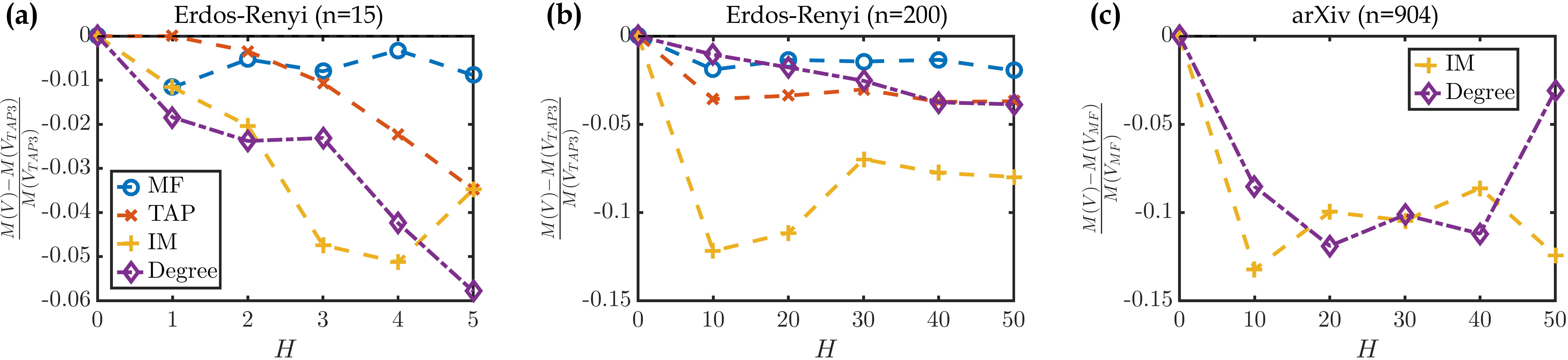}
\caption{\label{IsingComp} Comparison of the total Ising activity for greedy algorithms using various orders of the Plefka expansion. For each network, we ensure $\sum_j J_{ij} \le 1/2$ and we average over many draws of the initial bias $\{b_i^0\}\sim \mathcal{U}[-1/2,1/2]$. (a) An Erd\"{o}s-R\'{e}nyi network with $n=15$ nodes. The total activity is calculated exactly using the Boltzmann distribution. (b) An Erd\"{o}s-R\'{e}nyi network with $n=200$ nodes. (c) A collaboration network of $n=904$ physicists on the arXiv. The total activities in (b-c) are estimated using Monte Carlo simulations. In (a-b) the benchmark is TAP3, while for (c) the benchmark is MF.}
\end{figure*}

\begin{figure*}[t]
\centering
\includegraphics[width = .9\textwidth]{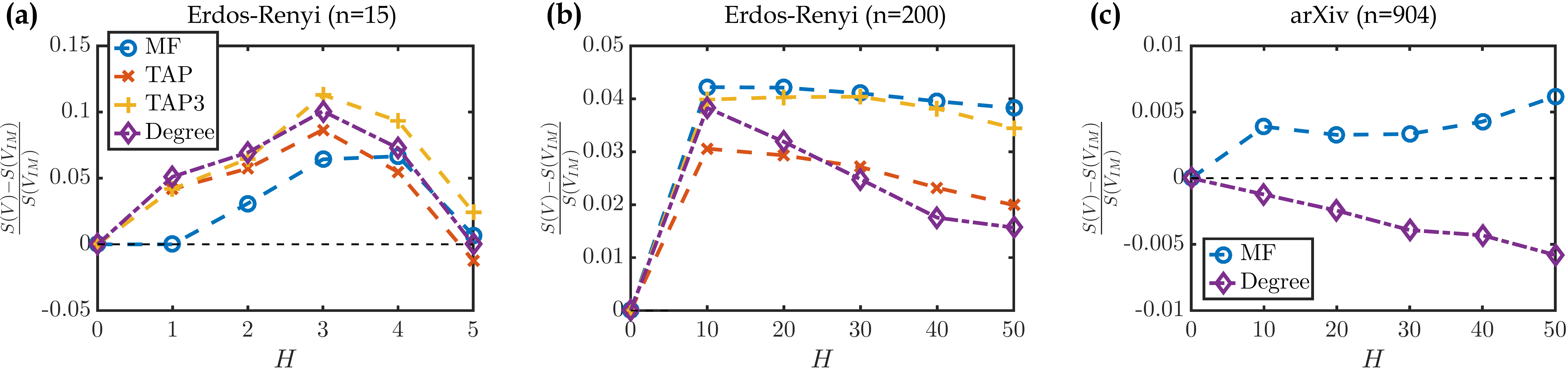}
\caption{\label{LTcomp} Comparison of the spread of influence under the linear threshold model for different greedy algorithms. For each network, we ensure $\sum_j J_{ij} \le 1/2$ and we average over many draws of the initial bias $\{b_i^0\}\sim \mathcal{U}[-1/2,1/2]$. (a) An Erd\"{o}s-R\'{e}nyi network with $n=15$ nodes. (b) An Erd\"{o}s-R\'{e}nyi network with $n=200$ nodes. (c) A collaboration network of $n=904$ physicists on the arXiv. The benchmark in all panels is IM.}
\end{figure*}

The greedy algorithm GI efficiently chooses an approximate set $V$ of influential nodes in $O(nH)$ iterations. However, $V$ could be arbitrarily far from the globally optimal set of nodes. Here, we present novel conditions for when $\mathcal{M}$ is monotonic and submodular in $V$, and, hence, GI is guaranteed to compute a $1-1/e$ approximation of the optimal set of nodes. The proof is based on the famous GHS inequality from statistical physics. \\

\noindent \textbf{Theorem 4 \cite{Griffiths-02}.} Given an Ising system with non-negative interactions $J\ge0$ and non-negative biases $\bm{b}\ge 0$, we have $\frac{\partial^2 \left<\sigma_i\right>}{\partial b_j\partial b_k} \le 0$ for all elements $i,j,k$. \\

We note that the GHS inequality guarantees a limited type of concavity of $\left<\sigma_i\right>$ in the direction of positive bias $\bm{b}$. While this was not enough to prove that PGA is optimal in the continuous setting, it is strong enough to guarantee that $\mathcal{M}$ is submodular in the discrete setting. \\

\noindent \textbf{Theorem 5.} For an Ising system with non-negative interactions $J\ge 0$ and non-negative initial biases $\bm{b}^{0}\ge 0$, the total activity $\mathcal{M}$ is monotonic and submodular in the activated nodes $V$. \\

\noindent \textit{Proof.} Monotonicity is guaranteed for any system with non-negative interactions in \cite{Griffiths-03}. To prove submodularity, we first introduce the notation $h_i^V \in \{0,1\}$ if $i$ is or is not in $V$. Then we note that $\mathcal{M}(V) \equiv\lim_{c\rightarrow\infty} M(\bm{b}^0 + c\bm{h}^V)$. Since $M$ is non-negative and concave in the direction of positive bias for $J\ge 0$ and $\bm{b}^0\ge 0$, $M$ it is subadditive. Thus, for any set $V$ of nodes and any two nodes $i,j\notin V$, we have
\begin{align}
&M(\bm{b}^0 + c(\bm{h}^V + \bm{h}^{\{i\}})) + M(\bm{b}^0 + c(\bm{h}^V + \bm{h}^{\{j\}})) \nonumber \\
&\ge M(\bm{b}^0 + c(\bm{h}^V + \bm{h}^{\{i\}} + \bm{h}^{\{j\}})) + M(\bm{b}^0 + c\bm{h}^V).
\end{align}
Taking $c\rightarrow\infty$, we find that
\begin{equation}
\mathcal{M}(V\cup \{i\}) + \mathcal{M}(V\cup \{j\}) \ge \mathcal{M}(V\cup \{i,j\}) + \mathcal{M}(V),
\end{equation}
which is the formal definition for submodularity.
\hfill $\square$

\subsection{Relationship Between the Linear Threshold and Ising Models}

It was recently established that the discrete version of the Ising influence maximization problem is closely related to traditional influence maximization in social networks \cite{Lynn-02}. In traditional influence maximization, one aims to maximize the spread of activations under a viral model, such as linear threshold (LT) or independent cascade. For example, the LT model is defined by the deterministic dynamics
\begin{equation}
\label{LT}
\sigma_i^{(t+1)} \leftarrow \sign\Big[\sum_j J_{ij}\sigma_j^{(t)} + b_i\Big].
\end{equation}
Typically, one considers activation variables $\sigma'_i\in \{0,1\}$ instead of $\sigma_i=\pm1$, which can be accomplished by a simple change of parameters $J'_{ij}\leftarrow 2J_{ij}$ and $b'_i \leftarrow b_i - \sum_jJ_{ij}$. The negative bias $\theta_i = -b'_i$ is referred to as the threshold of $i$, representing the amount of influence $i$ must receive from her neighbors to become active.

We include a stochastic influence $\epsilon$ for each node at every iteration $t$ of the LT dynamics, representing natural fluctuations in the influence on each node over time. If $\epsilon$ is drawn from a logistic distribution, then these stochastic dynamics are equivalent to Glauber Monte Carlo dynamics \cite{Newman-02}
\begin{equation}
P\big(\sigma_i^{(t+1)}\big|\bm{\sigma}^{(t)}\big) = \Big(1+e^{-\frac{2}{T}(\sum_j J_{ij}\sigma_j^{(t)} + b_i)}\Big)^{-1},
\end{equation}
where $T$ parameterizes the variance of $\epsilon$. Allowing the system time to equilibrate, the statistics of the Glauber dynamics follow the Boltzmann distribution (noting that we have taken $T=1$ in Eq. (\ref{Boltzmann})), simulating an Ising model. Furthermore, it is clear to see that we recover the deterministic LT dynamics in the limit of zero fluctuations $T\rightarrow 0$. Thus, the Ising model represents a natural generalization of the LT model to settings with stochastic fluctuations in the influence. We emphasize that, because maximum entropy models have demonstrated tremendous ability in quantitatively describing a wide range of real-world systems \cite{Schneidman-01,Ganmor-01,Bialek-01,Phillips-01,Stein-01,Kapur-01}, understanding how these systems react to external influence is a significant endeavor in and of itself, and this goal should fundamentally be viewed as running adjacent to, as opposed to in conflict with, the existing viral influence maximization literature.

Finally, we note that most applications of the LT model to influence maximization impose the constraint $\sum_jJ'_{ij} \le 1$ ($\sum_jJ_{ij} \le 1/2$ in Ising notation) and assume that the bias $b'_i$ is drawn uniformly from $[0,1]$ ($b_i\sim\mathcal{U}[-1/2,1/2]$) at the beginning of each simulation. Randomly selecting $b_i$ at the beginning of each simulation is meant to represent uncertainty in the nodes' biases, which is fundamentally distinct from randomizing $b_i$ at \textit{each iteration} to represent natural fluctuations in the biases over time. Indeed, in the following experiments we include both sources of randomness, making our model equivalent to the so-called random-field Ising model, which shows similar behavior to a spin glass \cite{Nattermann-01}.

\subsection{Experimental Evaluation}

We experimentally evaluate the performance of our greedy algorithm under various orders of the Plefka expansion. We also provide the first comparison between Ising influence algorithms and the traditional greedy influence maximization algorithm in \cite{Kempe-01}. As is usually assumed in traditional influence maximization, we scale the interactions such that $\sum_j J_{ij} \le 1/2$. Furthermore, we draw the initial bias on each node from a uniform distribution $b_i\sim\mathcal{U}[-1/2,1/2]$ and average over many such draws.

To fairly compare the Ising and linear threshold algorithms, we divide the experiments into two classes: the first is evaluated with respect to the total activity $\mathcal{M}$ under the Ising model, while the second class of experiments evaluates the spread $S$ resulting from each choice of nodes under the linear threshold model. We denote the greedy algorithm in \cite{Kempe-01} by IM. We also compare with the heuristic strategy of choosing the top $H$ nodes with the highest degrees in the network.

\subsubsection{Comparison Under the Ising Model.}

We first compare the different greedy algorithms under the Ising model. In Figs. \ref{IsingComp}(a-b), we use the Ising greedy algorithm GI with the third-order approximation TAP3 as the benchmark. In both Erd\"{o}s-R\'{e}nyi networks, we find that GI with TAP3 slightly outperforms TAP and MF, while all three Ising-based algorithms significantly out-perform the linear-threshold-based algorithm IM and the degree heuristic. Since TAP3, TAP, and MF all perform within $5\%$ of one another, in Fig. \ref{IsingComp}(c) we use GI with the MF approximation as the benchmark. In the arXiv network, we find that GI with MF significantly outperforms both LT and the degree heuristic. These results demonstrate the practical importance of accurately modeling the stochastic noise in the system. We point out that the total activity $\mathcal{M}$ is calculated exactly in Fig. \ref{IsingComp}(a) using the Boltzmann distribution and estimated in Figs. \ref{IsingComp}(b-c) using Monte Carlo simulations.

\subsubsection{Comparison Under the Linear Threshold Model.}

We now compare the algorithms under the linear threshold model. In all of Figs. \ref{LTcomp}(a-c), we use the LT greedy algorithm IM as the benchmark and exactly compute the spread of influence under the LT model. Surprisingly, in both of the Erd\"{o}s-R\'{e}nyi networks in Figs. \ref{LTcomp}(a-b), all of the Ising-based algorithms and the degree heuristic out-perform IM. In particular, the third-order approximation TAP3 achieves close to the largest spread in both networks. In the arXiv network in Fig. \ref{LTcomp}(c), the Ising-based algorithm continues to slightly out-perform IM, while IM out-performs the degree heuristic.

The success of the Ising-based algorithms is surprising, since they are all attempting to maximize a fundamentally different objective from influence spread under LT. We suspect that the strong performance might be the result of the Ising model performing a type of simulated annealing, similar to recent techniques proposed in \cite{Jiang-01}; however, an investigation of this hypothesis is beyond the scope of the current paper.

\section{Conclusions}

Maximum entropy models such as the Ising model have recently been used to quantitatively describe a plethora of biological, physical, and social systems. Given the success of the Ising model in capturing real-world systems, including populations of neurons in the brain and networks of interacting humans, understanding how to control and optimize the large-scale behavior of complex systems is of fundamental practical and scientific interest, with applications from guiding healthy brain development \cite{Goddard-01} and intervening to alleviate diseased brain states \cite{Goddard-02} to anticipating trends in financial markets \cite{Mantegna-01} and preventing viral epidemics \cite{Pastor-01}.

Here, we study the problem of maximizing the total activity of an Ising network given a budget of external influence. In the continuous setting where one can tune the influence on each node, we present a series of increasingly-accurate gradient ascent algorithms based on an approximation technique known as the Plefka expansion. In the discrete setting where one chooses a set of influential nodes, we propose a greedy algorithm and present novel conditions for when the objective is submodular.

\subsubsection{Future Work.} Given the novelty of this problem and the recent surge in the use of the Ising model, there are many promising directions to pursue. One direction is to consider a more general control problem in which the controller aims to shift the Ising network into a desired state instead of simply maximizing the activity of all nodes. 

Another direction is to consider data-based optimization, wherein the optimizer is only aware of the past activity of the system \cite{Goyal-02}. Since the Ising model is mathematically equivalent to a Boltzmann machine \cite{Coughlin-01}, one could adapt state-of-the-art methods from machine learning to approach this problem. 

Finally, given the experimental success of the Ising-based greedy algorithms under the linear threshold model, an obvious extension of the current work should look into possible explanations. We suspect that a closer comparison with simulated annealing techniques in \cite{Jiang-01} might provide the answer.

\section{Acknowledgements}

We thank Haim Sompolinsky and Eric Horsley for their helpful discussions. We acknowledge support from the U.S. National Science Foundation, the Air Force Office of Scientific Research, and the Department of Transportation.

%References and End of Paper
%These lines must be placed at the end of your paper
\bibliography{IsingInfluenceBib.bib}

\begin{thebibliography}{}

\bibitem[\protect\citeauthoryear{Ackley, Hinton, and
  Sejnowski}{1985}]{Ackley-01}
Ackley, D.~H.; Hinton, G.~E.; and Sejnowski, T.~J.
\newblock 1985.
\newblock A learning algorithm for boltzmann machines.
\newblock {\em Cog. Sci.} 9(1):147--169.

\bibitem[\protect\citeauthoryear{Aurell and Ekeberg}{2012}]{Aurell-01}
Aurell, E., and Ekeberg, M.
\newblock 2012.
\newblock Inverse ising inference using all the data.
\newblock {\em Phys. Rev. Lett.} 108(9):090201.

\bibitem[\protect\citeauthoryear{Bialek \bgroup et al\mbox.\egroup
  }{2012}]{Bialek-01}
Bialek, W.; Cavagna, A.; Giardina, I.; Mora, T.; Silvestri, E.; Viale, M.; and
  Walczak, A.~M.
\newblock 2012.
\newblock Statistical mechanics for natural flocks of birds.
\newblock {\em PNAS} 109(13):4786--4791.

\bibitem[\protect\citeauthoryear{Bolthausen}{2014}]{Bolthausen-01}
Bolthausen, E.
\newblock 2014.
\newblock An iterative construction of solutions of the tap equations for the
  sherrington--kirkpatrick model.
\newblock {\em Comm. Math. Phys.} 325(1):333--366.

\bibitem[\protect\citeauthoryear{Brush}{1967}]{Brush-01}
Brush, S.~G.
\newblock 1967.
\newblock History of the lenz-ising model.
\newblock {\em Rev. Mod. Phys.} 39(4):883.

\bibitem[\protect\citeauthoryear{Chen, Wang, and Wang}{2010}]{Chen-03}
Chen, W.; Wang, C.; and Wang, Y.
\newblock 2010.
\newblock Scalable influence maximization for prevalent viral marketing in
  large-scale social networks.
\newblock In {\em SIGKDD},  1029--1038.
\newblock ACM.

\bibitem[\protect\citeauthoryear{Chen, Wang, and Yang}{2009}]{Chen-02}
Chen, W.; Wang, Y.; and Yang, S.
\newblock 2009.
\newblock Efficient influence maximization in social networks.
\newblock In {\em SIGKDD},  199--208.
\newblock ACM.

\bibitem[\protect\citeauthoryear{Coughlin and Baran}{1995}]{Coughlin-01}
Coughlin, J.~P., and Baran, R.~H.
\newblock 1995.
\newblock {\em Neural computation in hopfield networks and boltzmann machines}.
\newblock University of Delaware Press.

\bibitem[\protect\citeauthoryear{Domingos and Richardson}{2001}]{Domingos-01}
Domingos, P., and Richardson, M.
\newblock 2001.
\newblock Mining the network value of customers.
\newblock In {\em KDD},  57--66.
\newblock ACM.

\bibitem[\protect\citeauthoryear{Duchi \bgroup et al\mbox.\egroup
  }{2008}]{Duchi-01}
Duchi, J.; Shalev-Shwartz, S.; Singer, Y.; and Chandra, T.
\newblock 2008.
\newblock Efficient projections onto the l 1-ball for learning in high
  dimensions.
\newblock {\em ICML}  272--279.

\bibitem[\protect\citeauthoryear{Galam}{2008}]{Galam-02}
Galam, S.
\newblock 2008.
\newblock Sociophysics: a review of galam models.
\newblock {\em Int. J. Mod. Phys. C} 19(3):409--440.

\bibitem[\protect\citeauthoryear{Ganmor, Segev, and
  Schneidman}{2011}]{Ganmor-01}
Ganmor, E.; Segev, R.; and Schneidman, E.
\newblock 2011.
\newblock Sparse low-order interaction network underlies a highly correlated
  and learnable neural population code.
\newblock {\em Proc. Natl. Acd. Sci.} 108(23):9679--9684.

\bibitem[\protect\citeauthoryear{Goddard, McIntyre, and
  Leech}{1969}]{Goddard-01}
Goddard, G.~V.; McIntyre, D.~C.; and Leech, C.~K.
\newblock 1969.
\newblock A permanent change in brain function resulting from daily electrical
  stimulation.
\newblock {\em Exp. Neurol.} 25(3):295--330.

\bibitem[\protect\citeauthoryear{Goddard}{1967}]{Goddard-02}
Goddard, G.~V.
\newblock 1967.
\newblock Development of epileptic seizures through brain stimulation at low
  intensity.
\newblock {\em Nature} 214(5092):1020--1021.

\bibitem[\protect\citeauthoryear{Goyal, Bonchi, and
  Lakshmanan}{2011}]{Goyal-02}
Goyal, A.; Bonchi, F.; and Lakshmanan, L.~V.
\newblock 2011.
\newblock A data-based approach to social influence maximization.
\newblock {\em VLDB Endowment} 5(1):73--84.

\bibitem[\protect\citeauthoryear{Griffiths, Hurst, and
  Sherman}{1970}]{Griffiths-02}
Griffiths, R.; Hurst, C.; and Sherman, S.
\newblock 1970.
\newblock Concavity of magnetization of an ising ferromagnet in a positive
  external field.
\newblock {\em J. Math. Phys.} 11:790.

\bibitem[\protect\citeauthoryear{Griffiths}{1967}]{Griffiths-03}
Griffiths, R.~B.
\newblock 1967.
\newblock Correlations in ising ferromagnets. i.
\newblock {\em J. Math. Phys.} 8(3):478--483.

\bibitem[\protect\citeauthoryear{Hillis}{1988}]{Hillis-01}
Hillis, W.~D.
\newblock 1988.
\newblock Intelligence as an emergent behavior; or, the songs of eden.
\newblock {\em Daedalus}  175--189.

\bibitem[\protect\citeauthoryear{Jaynes}{1957}]{Jaynes-01}
Jaynes, E.~T.
\newblock 1957.
\newblock Information theory and statistical mechanics.
\newblock {\em Phys. Rev.} 106(4):620.

\bibitem[\protect\citeauthoryear{Jiang \bgroup et al\mbox.\egroup
  }{2011}]{Jiang-01}
Jiang, Q.; Song, G.; Cong, G.; Wang, Y.; Si, W.; and Xie, K.
\newblock 2011.
\newblock Simulated annealing based influence maximization in social networks.
\newblock In {\em AAAI}, volume~11,  127--132.

\bibitem[\protect\citeauthoryear{Kapur}{1989}]{Kapur-01}
Kapur, J.~N.
\newblock 1989.
\newblock {\em Maximum-entropy models in science and engineering}.
\newblock John Wiley \& Sons.

\bibitem[\protect\citeauthoryear{Kempe, Kleinberg, and Tardos}{2003}]{Kempe-01}
Kempe, D.; Kleinberg, J.~M.; and Tardos, E.
\newblock 2003.
\newblock Maximizing the spread of influence through a social network.
\newblock In {\em KDD},  137--146.
\newblock ACM.

\bibitem[\protect\citeauthoryear{Leskovec \bgroup et al\mbox.\egroup
  }{2007}]{Leskovec-01}
Leskovec, J.; Krause, A.; Guestrin, C.; Faloutsos, C.; VanBriesen, J.; and
  Glance, N.
\newblock 2007.
\newblock Cost-effective outbreak detection in networks.
\newblock In {\em KDD},  420--429.
\newblock ACM.

\bibitem[\protect\citeauthoryear{L{\'e}vy}{1997}]{Levy-01}
L{\'e}vy, P.
\newblock 1997.
\newblock {\em Collective intelligence}.
\newblock Plenum/Harper Collins New York.

\bibitem[\protect\citeauthoryear{Lezon \bgroup et al\mbox.\egroup
  }{2006}]{Lezon-01}
Lezon, T.~R.; Banavar, J.~R.; Cieplak, M.; Maritan, A.; and Fedoroff, N.~V.
\newblock 2006.
\newblock Using the principle of entropy maximization to infer genetic
  interaction networks from gene expression patterns.
\newblock {\em PNAS} 103(50):19033--19038.

\bibitem[\protect\citeauthoryear{Lynn and Lee}{2016}]{Lynn-01}
Lynn, C.~W., and Lee, D.~D.
\newblock 2016.
\newblock Maximizing influence in an ising network: A mean-field optimal
  solution.
\newblock {\em NIPS}.

\bibitem[\protect\citeauthoryear{Lynn and Lee}{2017}]{Lynn-02}
Lynn, C.~W., and Lee, D.~D.
\newblock 2017.
\newblock Statistical mechanics of influence maximization with thermal noise.
\newblock {\em EPL}.

\bibitem[\protect\citeauthoryear{Lynn \bgroup et al\mbox.\egroup
  }{2017}]{Lynn-03}
Lynn, C.~W.; Papadopoulos, L.; Lee, D.~D.; and Bassett, D.~S.
\newblock 2017.
\newblock Surges of collective human activity emerge from simple pairwise
  correlations.
\newblock {\em Preprint}.

\bibitem[\protect\citeauthoryear{Mantegna and Stanley}{1999}]{Mantegna-01}
Mantegna, R.~N., and Stanley, H.~E.
\newblock 1999.
\newblock {\em Introduction to econophysics: correlations and complexity in
  finance}.
\newblock Cambridge University Press.

\bibitem[\protect\citeauthoryear{Mataric}{1993}]{Mataric-01}
Mataric, M.~J.
\newblock 1993.
\newblock Designing emergent behaviors: From local interactions to collective
  intelligence.
\newblock In {\em SAB},  432--441.

\bibitem[\protect\citeauthoryear{Mora \bgroup et al\mbox.\egroup
  }{2010}]{Mora-01}
Mora, T.; Walczak, A.~M.; Bialek, W.; and Callan, C.~G.
\newblock 2010.
\newblock Maximum entropy models for antibody diversity.
\newblock {\em PNAS} 107(12):5405--5410.

\bibitem[\protect\citeauthoryear{Namatame, Kurihara, and
  Nakashima}{2007}]{Namatame-01}
Namatame, A.; Kurihara, S.; and Nakashima, H.
\newblock 2007.
\newblock {\em Emergent Intelligence of Networked Agents}, volume~56.
\newblock Springer.

\bibitem[\protect\citeauthoryear{Nattermann}{1997}]{Nattermann-01}
Nattermann, T.
\newblock 1997.
\newblock Theory of the random field ising model.
\newblock {\em Spin glasses and random fields} 12:277.

\bibitem[\protect\citeauthoryear{Nesterov}{2013}]{Nesterov-01}
Nesterov, Y.
\newblock 2013.
\newblock {\em Introductory lectures on convex optimization: A basic course},
  volume~87.

\bibitem[\protect\citeauthoryear{Newman and Barkema}{1999}]{Newman-02}
Newman, M., and Barkema, G.
\newblock 1999.
\newblock {\em Monte Carlo Methods in Statistical Physics}.
\newblock New York: Oxford University Press.

\bibitem[\protect\citeauthoryear{Pastor-Satorras and
  Vespignani}{2001}]{Pastor-01}
Pastor-Satorras, R., and Vespignani, A.
\newblock 2001.
\newblock Epidemic spreading in scale-free networks.
\newblock {\em Phys. Rev. Lett.} 86(14):3200.

\bibitem[\protect\citeauthoryear{Phillips, Anderson, and
  Schapire}{2006}]{Phillips-01}
Phillips, S.~J.; Anderson, R.~P.; and Schapire, R.~E.
\newblock 2006.
\newblock Maximum entropy modeling of species geographic distributions.
\newblock {\em Ecol. Modell.} 190(3):231--259.

\bibitem[\protect\citeauthoryear{Plefka}{1982}]{Plefka-01}
Plefka, T.
\newblock 1982.
\newblock Convergence condition of the tap equation for the infinite-ranged
  ising spin glass model.
\newblock {\em J. Phys. A} 15(6).

\bibitem[\protect\citeauthoryear{Schneidman \bgroup et al\mbox.\egroup
  }{2006}]{Schneidman-01}
Schneidman, E.; Berry, M.~J.; II, R.~S.; and Bialek, W.
\newblock 2006.
\newblock Weak pairwise correlations imply strongly correlated network states
  in a neural population.
\newblock {\em Nature} 440(7087):1007.

\bibitem[\protect\citeauthoryear{Stein, Marks, and Sander}{2015}]{Stein-01}
Stein, R.~R.; Marks, D.~S.; and Sander, C.
\newblock 2015.
\newblock Inferring pairwise interactions from biological data using
  maximum-entropy probability models.
\newblock {\em PLoS Comp. Bio.} 11(7).

\bibitem[\protect\citeauthoryear{Sylvester}{1983}]{Sylvester-01}
Sylvester, G.~S.
\newblock 1983.
\newblock Magnetization concavity in ferromagnets.
\newblock {\em J. Stat. Phys.} 33(1):91--98.

\bibitem[\protect\citeauthoryear{Yedidia}{2001}]{Yedidia-01}
Yedidia, J.
\newblock 2001.
\newblock {\em Advanced mean field methods: Theory and practice}.

\end{thebibliography}


\begin{thebibliography}{}

\bibitem[\protect\citeauthoryear{Galam}{2008}]{Galam-02}
Galam, S.
\newblock 2008.
\newblock Sociophysics: a review of galam models.
\newblock {\em Int. J. Mod. Phys. C} 19(3):409--440.

\bibitem[\protect\citeauthoryear{Ganmor, Segev, and
  Schneidman}{2011}]{Ganmor-01}
Ganmor, E.; Segev, R.; and Schneidman, E.
\newblock 2011.
\newblock Sparse low-order interaction network underlies a highly correlated
  and learnable neural population code.
\newblock {\em Proc. Natl. Acd. Sci.} 108(23):9679--9684.

\bibitem[\protect\citeauthoryear{Georges and Yedidia}{1991}]{Georges-01}
Georges, A., and Yedidia, J.~S.
\newblock 1991.
\newblock How to expand around mean-field theory using high-temperature
  expansions.
\newblock {\em Journal of Physics A: Mathematical and General} 24(9).

\bibitem[\protect\citeauthoryear{Lynn and Lee}{2016}]{Lynn-01}
Lynn, C.~W., and Lee, D.~D.
\newblock 2016.
\newblock Maximizing influence in an ising network: A mean-field optimal
  solution.
\newblock {\em NIPS}.

\bibitem[\protect\citeauthoryear{Lynn \bgroup et al\mbox.\egroup
  }{2017}]{Lynn-03}
Lynn, C.~W.; Papadopoulos, L.; Lee, D.~D.; and Bassett, D.~S.
\newblock 2017.
\newblock Surges of collective human activity emerge from simple pairwise
  correlations.
\newblock {\em Preprint}.

\bibitem[\protect\citeauthoryear{Plefka}{1982}]{Plefka-01}
Plefka, T.
\newblock 1982.
\newblock Convergence condition of the tap equation for the infinite-ranged
  ising spin glass model.
\newblock {\em J. Phys. A} 15(6).

\bibitem[\protect\citeauthoryear{Schneidman \bgroup et al\mbox.\egroup
  }{2006}]{Schneidman-01}
Schneidman, E.; Berry, M.~J.; II, R.~S.; and Bialek, W.
\newblock 2006.
\newblock Weak pairwise correlations imply strongly correlated network states
  in a neural population.
\newblock {\em Nature} 440(7087):1007.

\bibitem[\protect\citeauthoryear{Thouless, Anderson, and
  Palmer}{1977}]{Thouless-01}
Thouless, D.~J.; Anderson, P.~W.; and Palmer, R.~G.
\newblock 1977.
\newblock Solution of 'solvable model of a spin glass'.
\newblock {\em Philosophical Magazine} 35(3):593--601.

\bibitem[\protect\citeauthoryear{Yedidia}{2001}]{Yedidia-01}
Yedidia, J.
\newblock 2001.
\newblock {\em Advanced mean field methods: Theory and practice}.

\bibitem[\protect\citeauthoryear{Zeng \bgroup et al\mbox.\egroup
  }{2011}]{Zeng-01}
Zeng, H.~L.; Aurell, E.; Alava, M.; and Mahmoudi, H.
\newblock 2011.
\newblock Network inference using asynchronously updated kinetic ising model.
\newblock {\em Physical Review E} 83(4):041135.

\end{thebibliography}
\bibliographystyle{aaai}

\end{document}

\title{Supplementary Information: Maximizing Activity in Ising Networks via the TAP Approximation}
\author{Christopher W. Lynn \\
Department of Physics \& Astronomy \\
University of Pennsylvania \\
Philadelphia, PA 19104, USA \\
\And Daniel D. Lee \\
Department of Electrical \& Systems Engineering \\
University of Pennsylvania \\
Philadelphia, PA 19104, USA}

\maketitle

\section{Appendix A: The Plefka Expansion}

Many complex systems in the biological and social sciences have recently been shown to be quantitatively described by the Ising model from statistical physics \cite{Schneidman-01,Ganmor-01,Lynn-03,Galam-02}. The Ising model is defined by the Boltzmann distribution:
\begin{align}
\label{Boltzmann}
P(\bm{\sigma}) &= \frac{1}{Z}\exp\left(- H(\bm{\sigma})\right), \\
H(\bm{\sigma}) &= -\frac{1}{2}\sum_{ij}J_{ij}\sigma_i\sigma_j - \sum_i b_i\sigma_i, \text{and} \\
Z &= \sum_{\bm{\sigma}\in \{\pm 1\}^n} e^{-\beta H(\bm{\sigma})},
\end{align}
where $H(\bm{\sigma})$ is the energy function, or Hamiltonian, that defines the system, and $Z$ is a normalization constant. The parameters $J = \{J_{ij}\}$ define the network of interactions between elements and the parameters $\bm{b} = \{b_i\}$ represent individual biases, which can be altered by application of targeted external influence.

Since the Ising model has remained unsolved for all but a select number of special cases, tremendous interdisciplinary effort has focused on developing tractable approximation techniques. Here, we present an advanced approximation method known as the Plefka expansion \cite{Yedidia-01}. The Plefka expansion is not an approximation itself, but is rather a principled method for deriving a series of increasingly accurate approximations, the first two orders of which are the na\"{i}ve mean-field (MF) and Thouless-Anderson-Palmer (TAP) approximations.

The Plefka expansion is a small-interaction expansion derived by Georges and Yedidia in \cite{Georges-01}, extending the work of Thouless, Anderson, Palmer, and Plefka \cite{Thouless-01,Plefka-01}. Throughout this section, we assume that the interactions are symmetric (i.e., $J = J^T$) to ease the presentation; however, we note that the Plefka expansion easily generalizes to include asymmetric interactions as well \cite{Lynn-01,Zeng-01}. Furthermore, because we are expanding in the limit $J\ll 1$, it is useful to take $J\rightarrow \beta J$, where $\beta$ parameterizes the strength of interactions in the system. 

Given an Ising system defined by $J$, $\bm{b}$, and $\beta$, we consider the free energy $G = -\ln (Z)/\beta$, which can be thought of as a mathematical tool whose derivative with respect to the external field $b_i$ defines the average activity of node $i$; i.e., $\left<\sigma_i\right> = -\partial G/\partial b_i$.
To derive the Plefka expansion, we consider the approximate free energy:
\begin{equation}
\label{Gibbs}
\beta \tilde{G} = -\ln\sum_{\{\bm{\sigma}\}} \exp\Big(-\beta H(\bm{\sigma}) + \sum_i \lambda_i (\sigma_i-m_i)\Big),
\end{equation}
where $\lambda_i$ are auxiliary fields that impose the constraints $m_i=\left<\sigma_i\right>$ and are eventually set to zero to recover the true free energy. Using (\ref{Gibbs}), we can expand the free energy around the small-interaction limit $\beta=0$. Carrying out this expansion to third-order in $\beta$, we obtain \cite{Yedidia-01}:
\begin{align}
\label{Expansion}
\beta G = &\sum_i \frac{1+m_i}{2}\ln\left(\frac{1+m_i}{2}\right) + \frac{1-m_i}{2}\ln\left(\frac{1-m_i}{2}\right) \nonumber \\
&-\beta\Big(\frac{1}{2} \sum_{ij} J_{ij}m_im_j + \sum_i b_i m_i\Big) \nonumber \\
&-\frac{\beta^2}{4}\sum_{ij} J_{ij}^2(1-m_i^2)(1-m_j^2) \nonumber \\
&-\frac{\beta^3}{3}\sum_{ij} J_{ij}^3 m_i(1-m_i^2)m_j(1-m_j^2) \nonumber \\
&- \frac{\beta^3}{6}\sum_{ijk} J_{ij}J_{jk}J_{ki}(1-m_i^2)(1-m_j^2)(1-m_k^2) \nonumber \\
&+ \hdots.
\end{align}
The zeroth-order term in the expansion corresponds to the mean-field entropy and the first-order term is the mean-field energy. Thus, up to first-order in $\beta$, we recover the MF approximation. The second-order term is known as the Onsager reaction term, and its inclusion yields the TAP approximation \cite{Thouless-01}. Higher-order terms are systematic corrections first presented in \cite{Georges-01} and, in principle, can be carried out to arbitrary order.

The quantities $m_i$ approximate the average activities $\left<\sigma_i\right>$ and are defined by the stationary conditions $\partial \tilde{G}/\partial m_i=0$. Thus, differentiating (\ref{Expansion}) with respect to $m_i$, and only keeping terms to second-order in $\beta$, we arrive at the TAP self-consistency equations for the magnetizations $\bm{m}$:
\begin{align}
\label{TAP}
m_i &= \tanh\Big[b_i + \sum_j J_{ij}m_j - m_i\sum_j J_{ij}^2(1-m_j^2) \Big] \nonumber \\
&\triangleq f^{\text{TAP}}_i(\bm{m}).
\end{align}
Thus, for any order $\alpha$ of the Plefka expansion, the intractable problem of computing exact averages over the Glauber dynamics is replaced by the manageable task of computing a fixed point of the corresponding self-consistency map $\bm{m} = \bm{f}^{(\alpha)}(\bm{m})$.

\section{Appendix B: The Third-Order TAP Approximation}

We now derive the self-consistency equations and response function for the third-order approximation in the Plefka expansion. To third-order in $\beta$, we arrive at the self-consistency equations:
\begin{align}
\label{SC}
m_i &= \tanh\left[b_i + \sum_j J_{ij} m_j - m_i\sum_j J_{ij}^2(1-m_j^2) \right. \nonumber \\
&\quad \quad\quad\quad + \frac{2}{3}(1-3m_i^2)\sum_j J_{ij}^3 m_j(1-m_j^2) \nonumber \\
& \quad\quad\quad\quad \left. - m_i\sum_{jk} J_{ij}J_{jk} J_{ki} (1-m_j^2)(1-m_k^2)\right] \nonumber \\
&\triangleq f^{\text{TAP3}}_i(\bm{m}).
\end{align}

For algorithmic purposes, we are also interested in the response function $\tilde{\chi}^{(\alpha)}_{ij} = \frac{\partial m_i}{\partial b_j}$, which, for any order $\alpha$ of the Plefka expansion, takes the form:
\begin{equation}
\tilde{\chi}^{(\alpha)} = (I-Df^{(\alpha)})^{-1}A,
\end{equation}
where $I$ is the identity matrix, $Df^{(\alpha)}_{ij} = \frac{\partial f^{(\alpha)}_i}{\partial m_j}$ is the Jacobian of the mean-field map, and $A_{ij} = (1-m_i^2)\delta_{ij}$ is a diagonal matrix. Thus, the response function for each extended mean-field approximation is defined by the Jacobian of the corresponding mean-field map. Up to third-order in $\beta$, $Df^{\text{TAP3}}$ takes the form:
\begin{align}
\label{Jacobian}
Df^{\text{TAP3}}_{ij} = (1-m_i^2 &) \Bigg[ J_{ij}  + 2 J_{ij}^2m_im_j \nonumber \\
& - \delta_{ij}\sum_k J_{ik}^2(1-m_k^2) \nonumber \\
& - 4 m_i\delta_{ij}\sum_k J_{ik}^3 m_k(1-m_k^2) \nonumber \\
& + \frac{2}{3} J_{ij}^3(1-3m_i^2)(1-3m_j^2) \nonumber \\
& -  \delta_{ij} \sum_{k\ell} J_{ik}J_{k\ell}J_{\ell i} (1-m_k^2)(1-m_{\ell}^2) \nonumber \\
& + 4 J_{ij}m_i m_j \sum_k J_{jk}J_{ki} (1-m_k^2) \Bigg].
\end{align}

%References and End of Paper
%These lines must be placed at the end of your paper
\bibliography{IsingInfluenceBib.bib}
\bibliographystyle{aaai}